\documentclass[twocolumn,showpacs,amsmath,amssymb,pra]{revtex4}
\usepackage{graphicx}
\begin{document}
\title{Few-body bound state stability of dipolar molecules in two dimensions}
\author{A.~G. Volosniev, D.~V. Fedorov, A.~S. Jensen, and N.~T. Zinner}
\affiliation{Department of Physics and Astronomy, Aarhus University, DK-8000 Aarhus C, Denmark }
\date{\today}

\begin{abstract}
Bound structures among dipolar molecules in multilayers are 
a topic of great interest in the light of recent experiments that have
demonstrated the feasibility of the setup. While it is known that
two molecules in two adjacent layers will always bind, larger complexes have only
been scarcely addressed thus far. Here we prove rigorously that 
three- and four-body states will never be bound when the dipoles
are oriented perpendicular to the layers. The technique employed is 
general and can be used for more molecules/layers and other geometries.
Our analytical findings are supported by numerical calculations for 
both fermionic and bosonic molecules. Furthermore, we calculate the
reduction in intralayer repulsion necessary to bind large complexes and
estimate the influence of bound complexes in systems with many layers.
\end{abstract}

\maketitle

\section{Introduction.} 
The production and manipulation 
of ultracold molecular gases is a truly riveting physical adventure
at the moment \cite{dipexp}. 
Interacting dipolar molecules confined to two dimensions are currently
attracting substantial attention by way of its recent experimental 
realization and study of geometrical effects on chemical 
reactions \cite{miranda2011}. 
Theoretically, this reduced dimensional
setup has spurred immense interest and many promising proposals for 
novel many-body states exists \cite{review}. Many of 
these studies ignore or make rough guesses as to which few-body 
states are allowed in a setup with adjacent two-dimensional (2D) planes.
An understanding of the basic building blocks is obviously crucial 
when trying to understand the complicated correlations in the many-body
systems. The low-energy few-body spectrum with short-range interactions is 
well-known \cite{jen04} and gives rise to the famous Efimov spectrum 
in three dimensions (3D) \cite{ferlaino2010} and universal trimer states 
in 2D \cite{nie97,filipe2011}. However, the 
dipolar interactions is anisotropic and long-ranged and it is not clear
that similar structures emerge. 

\begin{figure}[ht!]
\includegraphics[scale=0.33]{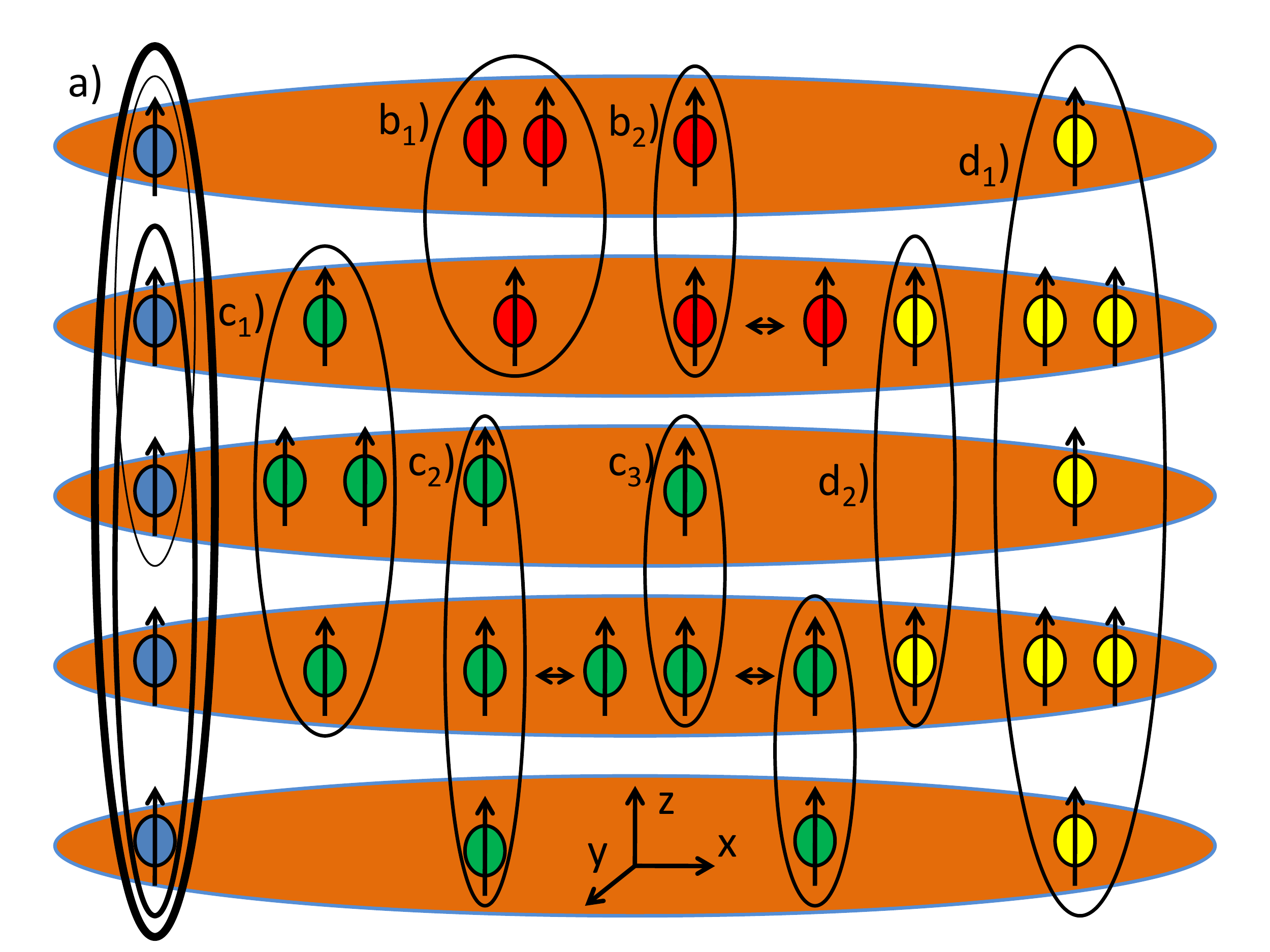}
\caption{(color online) Dipolar molecules in a multi-layer structure with dipoles oriented
perpendicular to the planes. $a$ illustrates the different chains that are the
most stable configurations. Three-body states are shown in $b_1$ and $b_2$, and
four-body states in $c_1$, $c_2$, and $c_3$. $d_1$ shows a non-chain five layer
configuration, while $d_2$ is a picture of a weakly bound dimer with twice
the layer separation.}
\label{fig1}
\end{figure}

A prominent example of the influence of bound states on the many-body
physics is the BCS-BEC crossover in a two-component Fermi gas with 
attractive short-range interactions \cite{bcsbec}. In the 2D case, 
the presence of a two-body bound dimer state plays a crucial role in the crossover
\cite{randeria1989}. This remains true for dipolar molecules in 
a bilayer with perpendicular polarization \cite{pikzinner2010}, 
and an effective
dimerization of the system along these lines have been suggested in the 
multilayer setup \cite{potter2010}. In the bilayer case, this can be infered
from the dipolar potential which always supports a bound two-body 
state \cite{shih2009,san10,vol11a,zinner11a}.
However, there is a gap in our 
current knowledge of the stability of few-body states containing more
than two molecules. In the present letter we give a complete
characterization of the bound state structures for up to five layers with 
perpendicularly polarized dipoles (see Fig.~\ref{fig1}).

The most stable three-body trimer system is a chain with one molecule in each
layer where the pairwise interactions are attractive at 
short distances \cite{arm11}.  Two
molecules in the same layer will repel each other.  The most interesting
trimer is therefore three molecules in two layers where two pairs can form
bound subsystems whereas one pair is entirely repulsive ($b_1$ in Fig.~\ref{fig1}).  
The question
is whether this trimer is bound with an energy below the dimer and 
a free molecule ($b_2$ in Fig.~\ref{fig1}).
This is a special case in 2D of a repelling pair where each is
attracted by a third particle.  The four-body tetramer system 
with a fourth
molecule in the layer next to the repelling pair ($c_1$ in Fig.~\ref{fig1})
is much closer to
stability with five bound two-body subsystems and only one
repelling pair. The question here is if that structure has a lower
energy than a bound string of three molecules and a free molecule ($c_2$ in Fig.~\ref{fig1}) 
where only three bound subsystems remain. This of course implies that
adding more molecules in favorable configurations could produce stable structures.

In the present letter we first rigourously prove that trimer and 
tetramer states are unstable for identical bosons or fermions in 
the layers. The mathematical method is very general and reduces the 
problem to a matter of geometrically identifying the threshold in 
the potential landscape of the systems. It can be applied to other
systems where repulsive and attractive terms compete, like the 
study of excitons in carbon nanotubes \cite{carbon,charged} or in organic semiconductors
\cite{micro,organic}. A suitable reduction of the in-plane dipolar repulsion 
leads to bound complexes as all interlayer interactions support bound
states. We expect that this can be achieved in different ways, 
either by applying an electric field gradient perpendicular to the planes, 
by having different molecules in the system, or by direct manipulation
of the dipole interaction in each layer through microwave transitions 
\cite{demille2002,micheli2007}.  Here we determine the critical repulsive
strengths where stable structures emerge analytically and support our
findings by detailed numerical calculations using the stochastic variational
method \cite{vol11b}.

\section{Formulation.}
The Schr\"{o}dinger equation for $N$ particles with masses $m$ is
\begin{equation} \label{e30}
   \bigg( - \sum_{i=1}^{N} \frac{\hbar^2}{2m}{\vec \nabla^2_{i} }
 + \sum_{i<j}^{N} \lambda_{ij} V_{ij}(|\vec r_i - \vec r_j|)\bigg) 
 \Psi = E_N \Psi \; ,
\end{equation}
with coordinates $\vec r_i$ and energy $E_N$.
The interaction between particles
$i$ and $j$ is $\lambda_{ij} V_{ij}$, i.e.
\begin{eqnarray} 
V_{ij}(r,n)= 
 \frac{\hbar^2 d}{m} 
 \frac{r^2 - 2n^2d^2}{(r^2 + n^2d^2)^{5/2}}  \;,
 \lambda_{ij}=m \frac{D_iD_j}{\hbar^2d} \;, \label{e34}
\end{eqnarray}
where $d$ is the interlayer distance and $D_i$ is the dipole moment
of molecule $i$. For $n=0$ we get the intralayer two-body $1/r^3$ repulsion, 
while $n>0$ produces the interlayer potentials when the separation is
$nd$. The strengths are defined by the product $D_iD_j$, 
and given through the dimensionless quantities
$\lambda_{ij}$. The $V_{ij}(r,n=0)$ is modified at small distance
by the transverse motion. This leads to the expression \cite{cremon2010}
\begin{eqnarray}
 V_{ij}(r,n=0) = \frac{\hbar^2 d}{m} \frac{1}{2\sqrt{2}l^3} 
 U(\frac{3}{2},1,\frac{r^2}{2l^2} ),
\end{eqnarray}
where $U$ is the confluent hypergeometric Kummer function and $l$ is the width 
of the layer. We have checked that our results are insensitive to $l$ in the interval
$0.1\leq l/d\leq0.2$ and the detailed properties at small distance are not important.
Note that we assume that there is no tunneling between the layers, which implies that 
symmetrization is only relevant for particles in the same plane.

\subsection{Proof of instability.}
Consider first the three-body system with molecules 1 and 2
in the same plane and molecule 3 in an adjacent plane ($b_1$ in Fig.~\ref{fig1}).  
We assume that $D_1=D_2$ which may be different from $D_3$.  We
denote the attractive and repulsive strengths by $\lambda_a \equiv
\lambda_{13} = \lambda_{23}$ and $\lambda_r \equiv \lambda_{12}$. 
Notice that both $\lambda_a$ and $\lambda_r$ are positive. The 
reason we talk of an attractive interaction in this case is 
that the potential always supports a bound state \cite{vol11a}.
We shall
now prove that the system is unbound for $\lambda_r=\lambda_a$.  
Clearly, the system is bound for a sufficiently small ratio
$\lambda_r/\lambda_a$. We can then, for any given $\lambda_a$, define
a critical repulsive strength, $\lambda_r^{cr}[\lambda_a]$ (which is 
thus a function of $\lambda_a$ as indicated by the brackets), as the
value where the energy precisely equals the two-body
energy, $E_2[\lambda_a]$, of the bound 13 and 23 (interlayer) subsystems,
i.e.
\begin{equation}
\lambda_r^{cr}[\lambda_a]=\frac{E_2[\lambda_a]-\langle\Psi|T+\lambda_a 
 V_{13}+\lambda_a V_{23}|\Psi\rangle} {\langle\Psi|V_{12}|\Psi\rangle} , \;\label{ins-1}
\end{equation}
where $T$ is the kinetic energy operator and $\Psi$ is the normalized
wave function at the threshold for binding.  We now consider 
$F_{cr}=\tfrac{\partial}{\partial\lambda_a}\left(\tfrac{\lambda_{r}^{cr}[\lambda_a]}{\lambda_a}\right)$.
We now make repeated use of the general fact that, for a Hamiltonian $h$ with 
eigenstate $|\phi\rangle$ and eigenvalue $\epsilon$, i.e.
$h|\phi\rangle= \epsilon |\phi\rangle $, the partial derivative with respect to 
paramters that enter $h$ (and therefore also enter $\epsilon$) is $\epsilon'=
\langle\phi|h'|\phi\rangle$. Applying this fact for the parameter $\lambda_a$, we obtain
\begin{equation}
\frac{\partial \lambda_r^{cr}[\lambda_a]}{\partial \lambda_a}=
\frac{\frac{\partial E_2[\lambda_a]}{\partial \lambda_a}-
\langle\Psi|V_{13}+V_{23}|\Psi\rangle}{\langle\Psi|V_{12}|\Psi\rangle}.\;\label{ins-2}
\end{equation}
Defining the two-body bound state wave function, $\Phi$, and the
related kinetic energy operator, $T_2$, we get
\begin{eqnarray}
 \lambda_a^2 F_{cr} &=& 
 \frac{\lambda_a\frac{\partial E_2[\lambda_a]}{\partial \lambda_a}-E_2[\lambda_a]
 +\langle\Psi|T|\Psi\rangle}{\langle\Psi|V_{12}|\Psi\rangle} \nonumber  \\  \label{ins-4} &=&  
\frac{\langle\Psi|T|\Psi\rangle-\langle\Phi|T_2|\Phi\rangle}{\langle\Psi|V_{12}|\Psi\rangle}. 
\end{eqnarray}
The structure of $\Psi$ is the product, $\Psi=\psi \Phi$, where $\psi$
only depends on the relative coordinate, ${\bf y}$, between the dimer, $\Phi$,
and the third molecule.  This follows from the $r^{-3}$ tail of the dipolar potential
which is of shorter range than $r^{-2}$, which prohibits
long-range correlations \cite{jen04}. The universal limit with a 
two-body cluster structure is therefore approached \cite{arm10a}.  
We immediately conclude that
$\langle\Psi|T|\Psi\rangle-\langle\Phi|T_2|\Phi\rangle = \langle\psi|T_y|\psi\rangle $ is positive, where
$T_y=T-T_2$ is the kinetic energy operator depending only on
${\bf y}$. Thus, $F_cr>0$ and the relative critical strength increases with $\lambda_a$.

We now compute $\lambda_r^{cr}[\lambda_a]/\lambda_a$ for $\lambda_a = \infty$,
which is equivalent to infinite masses and negligibly small kinetic
energies.  The critical repulsive strength is found when the
total potential in all space becomes larger than the
two-body energy. Thus, the potential surface,
$V_{13}+V_{23}+V_{12}\lambda_r^{cr}[\lambda_a]/\lambda_a$, depending on two
relative coordinates, should have a minimum less than the minimum of
the two-body potential in Eq.~\eqref{e34}. This geometrical condition yields
$\lambda_r^{cr}[\lambda_a]/\lambda_a \le 0.373$. Since $F_{cr}>0$, this is a strict
inequality for all finite values of $\lambda_a$. In particular, 
the system is unbound for all $\lambda_a$ when $\lambda_r=\lambda_a$.

The most stable four-body system is a chain with one molecule per
layer (subset of $a$ in Fig.~\ref{fig1}), followed by the $c_1$ configuration 
in Fig.~\ref{fig1}. Four molecules in two layers are much less stable 
due to the additional repulsion.
To study the stability when changing the repulsion, we consider two
molecules in a single layer with $D_2=D_3$ and one molecules in 
each of the layers above and below it with $D_1=D_4$, keeping the 
product $D_1 D_3$ fixed.
The strengths are $\lambda_{a} = \lambda_{13} =
\lambda_{12} =\lambda_{24} =\lambda_{34}$, $\lambda_{r} = \lambda_{23}$, 
and $\lambda_{14} =\lambda_{a}^2 /\lambda_{r}$.  
The
threshold structure, $\Psi$, approached when $\lambda_{r} \rightarrow
\lambda_r^{cr}[\lambda_a]$, is three molecules in three layers, $\Phi$, and
a free fourth molecule ($c_2$ in Fig.~\ref{fig1}).  
The condition of equal energies
analogous to Eq.~\eqref{ins-1} becomes
\begin{eqnarray}
&\langle\Psi|T+\lambda_r^{cr}[\lambda_a]V_{23}+\frac{\lambda_a^2}{\lambda_r^{cr}[\lambda_a]}V_{14}+
\lambda_a(V_{13}+V_{12} +V_{24}  \label{ins-5}
+V_{34}) |\Psi\rangle& \nonumber \\&=\langle\Phi|T_3+\lambda_a V_{13} +\lambda_a V_{24}+
\frac{\lambda_a^2}{\lambda_r^{cr}[\lambda_a]}V_{14}|\Phi\rangle.&
\end{eqnarray} 
Similar to the three-body case, the threshold structure is
$\Psi = \psi \Phi$, where $\psi$ is the wave function
depending on the relative coordinate, ${\bf y}$, between the trimer
and the fourth molecule.
This means that
$\langle\Psi|V_{14}|\Psi\rangle=\langle\Phi|V_{14}|\Phi\rangle$, implying that the
critical value easily is found from Eq.~\eqref{ins-5}.  A completely
analogous procedure as above yields
\begin{equation}
 \lambda_a^2 F_{cr}
= \frac{\langle\Psi|T|\Psi\rangle-\langle\Phi|T_3|\Phi\rangle}{\langle\Psi|V_{23}|\Psi\rangle},
\label{ins-7}
\end{equation}
where $T_3$ is the three-body kinetic energy operator.
This derivative is postive and an upper bound for
stability is obtained from $\lambda_a= \infty $.  The geometries of the
potentials are such that $\lambda_r^{cr}[\lambda_a]/\lambda_a \le
0.75$, and all these four-body systems in two or three layers are unstable.

Our results strongly indicate that additional layers could lead to
stable configurations with more than one molecule in each layer. The
proofs above hold for bosons and distinguishable particles. However,
the threshold for fermions is always higher and produces less stable 
systems.

\begin{figure}[ht!]
\includegraphics[scale=0.75]{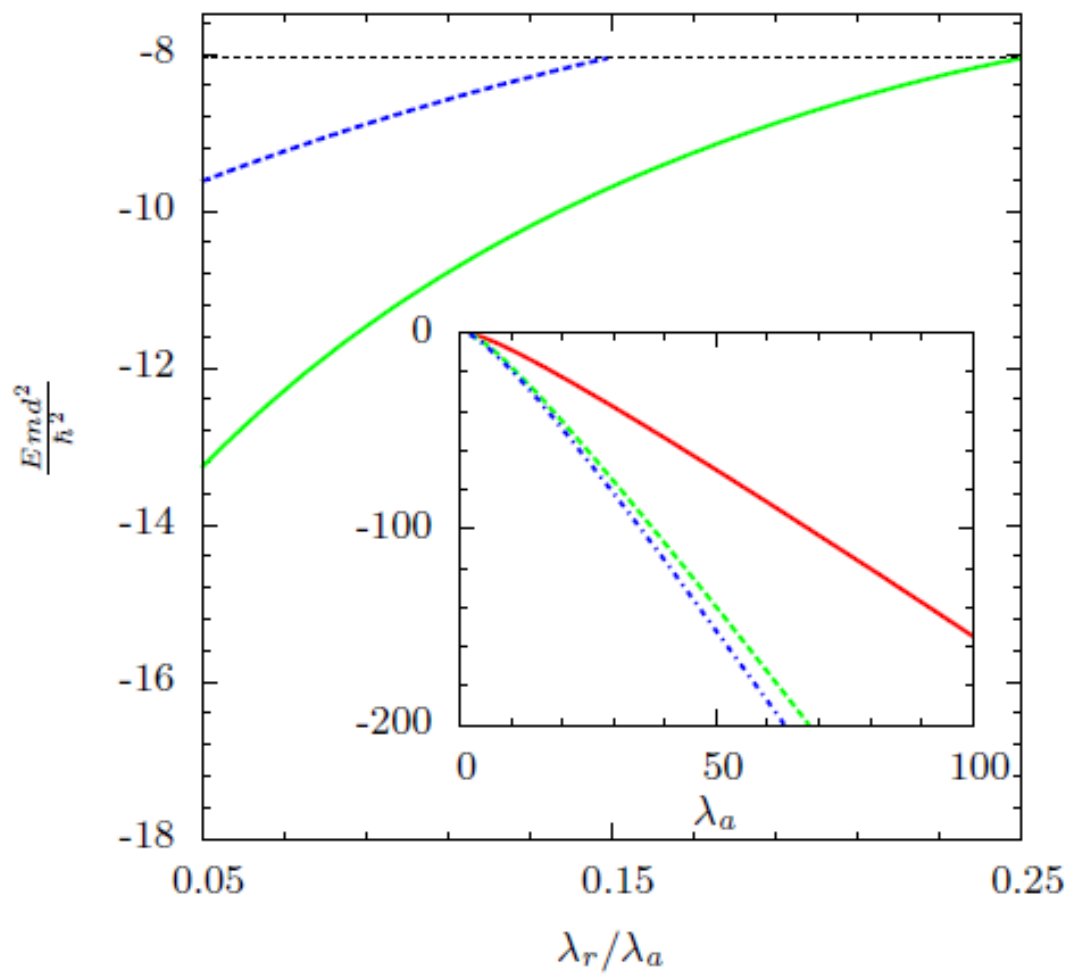}
\caption{(color online) Numerical trimer enegies as function of
  $\lambda_r/\lambda_a=D_1/D_3$ of the configuration $b_1$
  in Fig.~\ref{fig1} calculated with $\lambda_a=10$ for 
  bosons (solid green) and fermions (dashed blue). A horizontal
  dashed marks the two-body threshold. The inset compares
  the two-layer dimer (solid red) with the three-layer trimer as function 
  of $\lambda_a$ (dot-dashed blue). The (green) dashed curve 
  is twice the dimer energy demonstrating that the three-layer 
  trimer is more stable than two free dimers.}
\label{fig2}
\end{figure}

\section{Numerical results.}
\begin{figure}[ht!]
\includegraphics[scale=0.75]{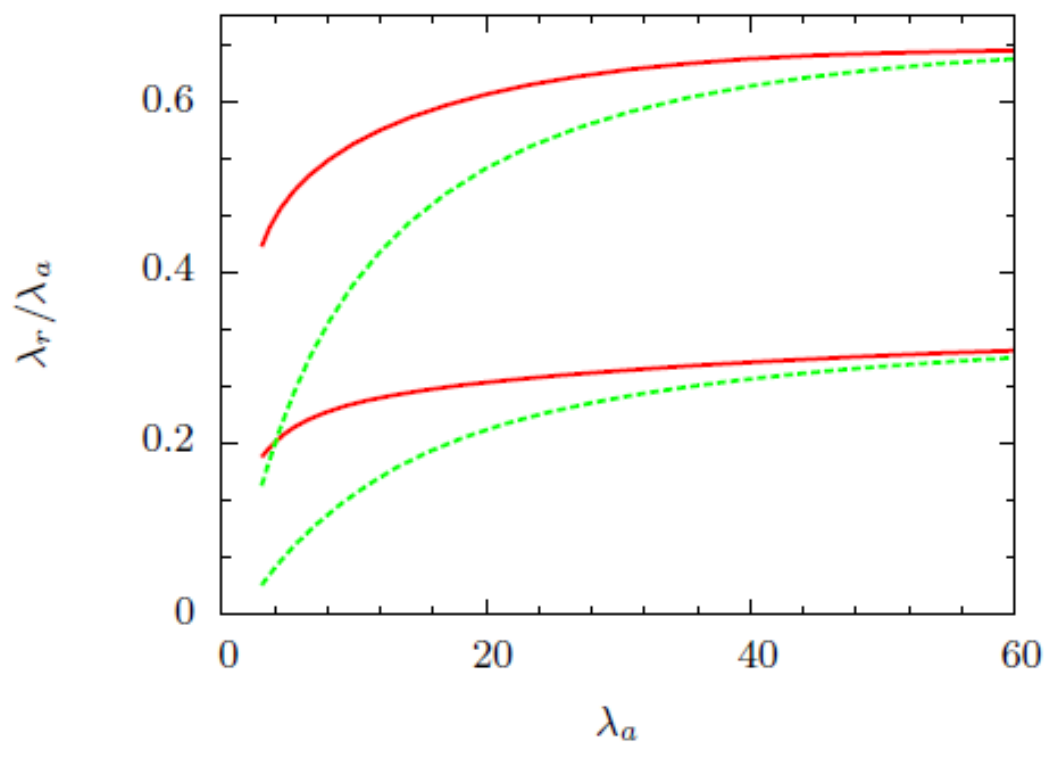}
\caption{(color online) The critical strengths, $\lambda^{cr}_r/\lambda_a$, as
  functions of $\lambda_a$, for configurations $b_1$ (lower curves)
  and $c_1$ (upper curves) of Fig.~\ref{fig1} with fermionic (solid)
  and bosonic (dashed) species.
  The asymptotic values, $0.373$ and $0.75$, for large $\lambda_a$ are
  slowly approached.}
\label{fig3}
\end{figure}

We illustrate our findings numerically using the stochastic 
variational method with correlated gaussians \cite{vol11b}. Fig.~\ref{fig2}
shows two- and three-body energies as function of $\lambda_r/\lambda_a$
with fixed $\lambda_a=10$ for the $b_1$ configuration in Fig.~\ref{fig1}.
The trimer energies increase with increasing repulsion, and become
unstable at the dimer threshold. 
Resonances could exist at higher
energies but the effective barriers are tiny for these potentials.  A
large width would immediately be accumulated and the resonance
disappears by merging into the two-body continuum background.
The fermion energies exceed the boson energies due to the antisymmetry 
which itself costs energy compared to symmetric systems.  This is
unavoidable in spite of the repulsion because the 
attraction at short distance must be larger to
provide a bound state, implying that the two molecules in the same
layer cannot be far apart. 
The distance from stability is reflected in the critical repulsion
below which the system is bound. The lower curves in Fig.~\ref{fig3}
show the numerical results for trimers. At large $\lambda_a$ the analytical 
results $0.373$ is approached very slowly, while at small $\lambda_a$ only a small 
repulsive perturbation is needed to break the exponentially weakly bound 
system.

\begin{figure}[ht!]
\includegraphics[scale=0.75]{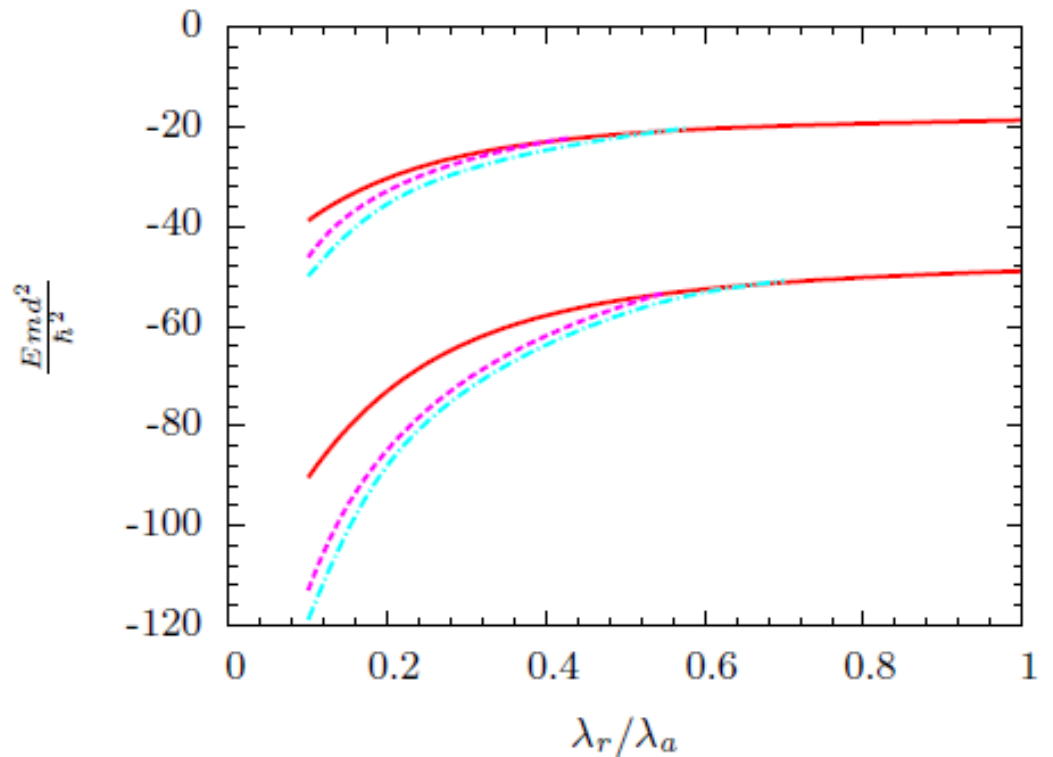}
\caption{(color online) Tetramer energies as function of $\lambda_r/\lambda_a$ in 
configuration $c_1$ of Fig.~\ref{fig1} with $\lambda_a=10$ (upper)
and $\lambda_a=20$ for fermions (dashed purple) and bosons (dot-dashed blue). 
The solid red curve shows the threshold energy ($c_2$ of Fig.~\ref{fig1}).}
\label{fig4}
\end{figure}

Consider now the four-body system in configuration $c_1$ of Fig.~\ref{fig1}.
As demonstrated in the inset of Fig.~\ref{fig2}, a chain of 
three molecules in three layers is more bound that two dimers. 
This can be understood since it gains energy from the attractive
interaction of configuration $d_2$ in Fig.~\ref{fig1}.
The threshold for binding $c_1$ is therefore a chain of 
three molecules and a free molecule, $c_2$, rather than $c_3$. The
energies for the $c_1$ system are shown in Fig.~\ref{fig4}
where we again see that the threshold is reached for $\lambda_r/\lambda_a<1$.
The relative critical strengths are shown in
Fig.~\ref{fig3}, where the asymptotic strong interaction limit of $0.75$
obtained from geometry is approached.

\subsection{Structure.}
The structure of trimer and tetramer is contained in relative
distances within the bound state. 
In the strong dipole limit, all molecules remain close
until the repulsion is extremely close to the threshold where one
molecule eventually leaks out through the repulsive barrier. The
universal limit of $\langle r^2\rangle |E|\to\tfrac{1}{3}$ is 
approached from below but only for very small energies \cite{arm10a,vol11a}.  
For weak
dipoles, the wave functions are already distributed over large regions
of space. Our numerical calculations confirm these expectations that 
are used in the proof of instability.

\section{Conclusions and Perspectives.}
We have demonstrated that in two or three layer system with 
perpendicularly polarized dipolar molecules, no three- or four-body 
bounds states are stable. This was shown rigorously through a 
novel analytical approach that predicts geometrically how 
much the repulsive interactions in the system must be reduced 
to stabilize such states. We also presented numerical results
in support of our conclusions and predicting exact critical 
values for when bound state with more than two molecules
are stable in the setup. Fermionic systems are found to 
be less stable than bosonic for all systems studied. Here we 
have studied the case of perpendicular polarization, but we note
that similar arguments can be applied for moderate tilting angles
where the in-plane interaction remains repulsive on average.

While we find that chains are the most stable 
configurations ($a$ in Fig.~\ref{fig1}), 
our results show that four-body states are much closer to stability
than three-body states.  This implies that an increase in the
number of layers can lead to bound complexes with more than 
one molecule per layer (as for instance shown in configuration $d_1$
of Fig.~\ref{fig1}). Disregarding these few-body states, both
density wave \cite{density} and crystalline \cite{crystal} phases have
been predicted in the multilayer setup.
With a richer spectrum of few-body complexes, we expect that 
more complicated, stable many-body
configurations are possible when the
complexes discussed here are used as building blocks.
Detecting the structures experimentally should be possible by 
several techniques such as RF spectroscopy \cite{shin2007}, 
lattice shaking \cite{strohmaier2010}, or {\it in situ} optical 
detection \cite{wunsch2011,wunsch2011b}. 

As demonstrated here, one can tune bound state stability by 
reducing the intra-layer repulsion. 
If an electric field gradient is applied that is steep enough to
make the interaction in adjacent layer different the 
effective repulsion could be reduced and stabilize few-body bound
states. Alternatively, one can imagine driving
microwave transitions among rotational
states in the molecules \cite{micheli2007,demille2002} into states
of different effective dipole moment for a given external dc field. 
Through this technique it might be possible to even reverse the direction
of the effective dipole moment, yielding a potential that is 
repulsive at short distance and attractive in the tail. This 
will produce entirely new structures and is an exciting direction for
future studies.

\acknowledgments
We thank J. Arlt, J.~F. Sherson, and B. Wunsch for enlightening discussions.

\end{document}